\documentclass[10pt,letterpaper,twocolumn]{article}
\usepackage{tipa} %% two column, final layout
\usepackage{ol2}
\usepackage[draft]{hyperref}
\usepackage{amsmath}
\usepackage{txfonts}
\begin{document}

\twocolumn[ %% activate for two-column option

\title{Generation of arbitrary cylindrical vector beams on
the higher order Poincar\'{e} sphere}

%% For REVTeX it is possible to automate superscript and e-mail callouts with the superscriptaddress option; see REVTeX4 documentation.

\author{Shizhen Chen, Xinxing Zhou, Yachao Liu, Xiaohui Ling,  Hailu Luo,$^{*}$ and Shuangchun Wen}
\address{
$^1$Key Laboratory for Micro-/Nano-Optoelectronic Devices of
Ministry of Education, College of Physics and Microelectronics
Science, Hunan
University, Changsha 410082, China\\
$^*$Corresponding author: hailuluo@hnu.edu.cn}

\begin{abstract}
We propose and experimentally demonstrate a novel interferometric
approach to generate arbitrary cylindrical vector beams on the
higher order Poincar\'{e} sphere. Our scheme is implemented by
collinear superposition of two orthogonal circular polarizations
with opposite topological charges. By modifying the amplitude and
phase factors of the two beams, respectively, any desired vector
beams on the higher order Poincar\'{e} sphere with high tunability
can be acquired. Our research provides a convenient way to evolve
the polarization states in any path on the high order Poincar\'{e}
sphere.

\end{abstract}

] %% activate for two-column option

\noindent  Light beam with spatially inhomogeneous state of
polarization, also referred to as vector beam, has been investigated
for many years due to its unique properties~\cite{Zhan2009}.
Comparing with the conventional homogeneous polarization represented
by fundamental Poincar\'{e} sphere, the cylindrical vector beams can
be represented by higher order Poincar\'{e} sphere
(HOPS)~\cite{Holleczek2010,Milione2011,Cardano2012}. Particular
interests and investigations focused on the vector beams with radial
and azimuthal polarizations, which can be represented as two points
on the equator of the first-order Poincar\'{e} sphere. Such beams
can be generated by twisted nematic liquid
crystal~\cite{Stadler1996,Marrucci2006,Han2013}, inserting phase
elements in the laser resonator~\cite{Oron2000}, computer-generated
subwavelength dielectric gratings~\cite{Bomzon2002,Levy2004}, a
conical Brewster prism~\cite{Kozawa2005}, spatially variable
retardation plates~\cite{Machavariani2007}, and a binary phase
mask~\cite{Ye2014}. The vector beams with special polarization
symmetry can give rise to uniquely high-numerical-aperture focusing
properties that may find important applications in nanoscale optical
imaging and
manipulation~\cite{Zhan2004,Zhan2006,Deng2007,Salamin2007,Ling2012,Yang2013}.

In this Letter, a novel interferometric method is proposed and
experimentally demonstrated to generate arbitrary cylindrical vector
beams on the HOPS. Homogeneous polarization on the fundamental
Poincar\'{e} sphere can be seen as the superposition of two
orthogonal circular polarizations corresponding to the two poles of
the Poincar\'{e} sphere. Similarly, vector beams on the HOPS can be
regarded as the linear superposition of two orthogonal circular
polarizations with opposite topological charges. For homogenous
polarization, two quarter-wave plates (QWPs) and one half-wave plate
(HWP) with adjustable optical axis angles can transform it to any
point on the fundamental Poincar\'{e} sphere~\cite{Holleczek2010}.
For the HOPS, we use a modified Mach-Zender interferometer with
which the amplitude and phase factors in each arm can be modified,
respectively.

\begin{figure}
\centerline{\includegraphics[width=8cm]{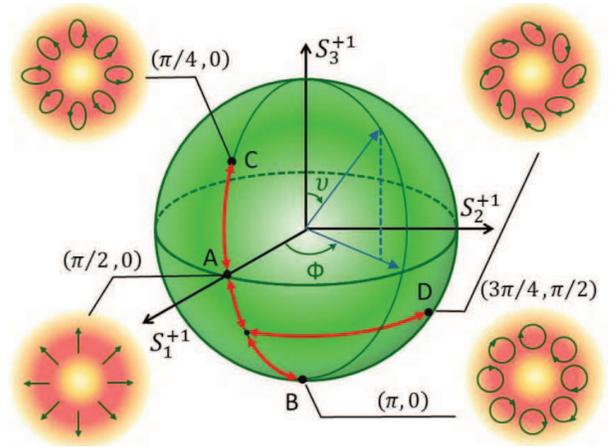}}
\caption{\label{Fig1} Schematic illustrating of the first-order
Poincar\'{e} sphere ($\ell=+1$). The points $A$, $B$, $C$, and $D$
represent four types of vector beams, respectively. The
transformations corresponding to the paths $AB$, $AC$, and $AD$ are
illustrated by the red curve with arrows. The insets show the
intensity distribution and the polarization vector (green arrows). }
\end{figure}

In the parameter space of the HOPS, the state of polarization
$\psi_{\ell}$ can be represented by~\cite{Milione2011}
\begin{equation}
\psi_{\ell}(\upsilon,\phi)=\cos\left(\frac{\upsilon}{2}\right)
e^{-i\phi/2}\mathrm{L}_{\ell}+\sin\left(\frac{\upsilon}{2}\right)
e^{i\phi/2}\mathrm{R}_{\ell}\label{AVP}.
\end{equation}
Here, $\phi$ is the azimuthal angle and $\upsilon$ the polar angle
in the spherical coordinate, respectively. $\mathrm{L}_{\ell}$ and
$\mathrm{R}_{\ell}$ are orthogonal circular polarization vortexes
with $\mathrm{L}_{\ell}=(\hat{x}+i\hat{y})
e^{-i\ell\varphi}/\sqrt{2}$ and
$\mathrm{R}_{\ell}=(\hat{x}-i\hat{y}) e^{i\ell\varphi}/\sqrt{2}$,
possessing spin angular momentum $\sigma\hbar$ ($\sigma=\pm{1}$) per
photon where $\hbar$ is the Plank constant. The factor
$e^{i\ell\varphi}$ is the vortex phase associated with the orbital
angular momentum $\ell\hbar$ per photons where $\ell$ is an integer
number ($\ell=\pm{1}, \pm{2}, \pm{3},...$)~\cite{Allen1992}.

For arbitrary points on the HOPS, the state of polarization
$\psi_{\ell}(\upsilon,\phi)$ can be described as the linear
combination of two orthogonal circular polarizations with opposite
topological charges. Equation~(\ref{AVP}) indicates that
$\cos(\upsilon/2)$ and $\sin(\upsilon/2)$ are amplitude factors,
while $\exp(-i\phi/2)$ and $\exp(i\phi/2)$ are phase factors,
respectively. By separately modifying the amplitude and phase
factors of the two orthogonal components, any desired vector beam on
the HOPS can be achieved. Generally, the equatorial points on the
HOPS represent linear polarized vector beams, the two poles denote
orthogonal circular polarizations with opposite topological charges,
and other points are elliptically polarization, as shown in
Fig.~\ref{Fig1}.

\begin{figure}
\centerline{\includegraphics[width=8cm]{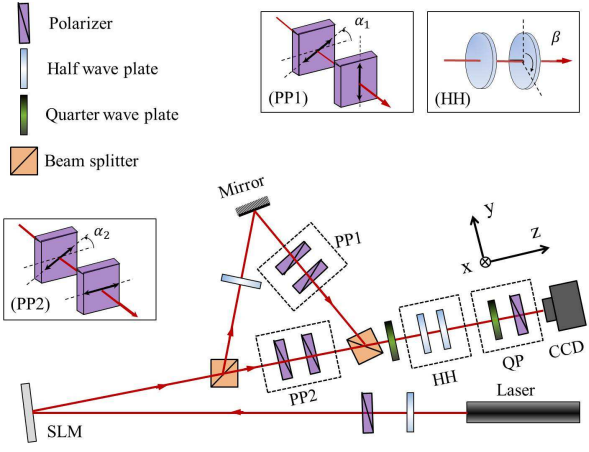}}
\caption{\label{Fig2} Experimental setup: Gaussian beam generated by
a He-Ne laser ($632.8\mathrm{nm}$, Thorlabs HNL210L-EC) passes
through a HWP and a Glan laser polarizer (GLP) to produce a
horizontally polarized beam. Then the fundamental Gaussian beam is
transformed into a vortex-bearing $\mathrm{LG}_{0}^{1}$ mode by a
spatial light modulator (SLM) (Holoeye Pluto-Vis). The modified
Mach-Zender interferometer comprising of two cascaded beam splitters
(BSs) and an odd number of reflections can convert the
$\mathrm{LG}_{0}^{1}$ mode into $\mathrm{LG}_{0}^{-1}$ mode. In
addition, we add PP1 and PP2 (two cascaded GLPs of which the second
one is fixed) in each arm of the interferometer to justify the
intensity factors. HH (a pair of HWPs within which only the second
one can be rotated) is added to modify the phase factors. QP
consisting of QWP and GLP, is used to measure Stokes parameters. The
intensity distribution of the field is detected by a CCD camera
(Coherent LaserCam HR).}
\end{figure}

Our experimental setup is plotted in Fig.~\ref{Fig2}. We use a
modified Mach-Zender interferometer, similar to that in
Ref.~\cite{Milione2012}, to generate the vector beams. In addition,
we employ two cascaded GLPs in each arm to modify the amplitude
factors and two cascaded HWPs to modify the phase factors,
respectively (see Fig.~\ref{Fig2}). Modulating the magnitude factor,
we can transform the polarization along the longitudes on the HOPS.
For northern hemisphere, the relation of $\upsilon$, $\alpha_{1}$,
and $\alpha_{2}$ can be written as
\begin{equation}
\alpha_{1}=90^{\circ},~~~\cos^{2}\alpha_{2}=\tan\frac{\upsilon}{2}
\label{HPT},
\end{equation}
and on southern hemisphere, it becomes
\begin{equation}
\sin^{2}\alpha_{1}=\cot\frac{\upsilon}{2},~~~
\alpha_{2}=0^{\circ}\label{VPT},
\end{equation}
where $\alpha_{1}$ and $\alpha_{2}$ are rotation angles with respect
to the $x$ axis as illustrated in the insets of Fig.~\ref{Fig2}. By
rotating the second HWP of the HH, we can adjust the phase factors
of the vector beams, thereby transforming the polarization along the
latitudes on the HOPS. The relation between $\phi$ and $\beta$ is
given by
\begin{equation}
\beta=\phi/4\label{centroid},
\end{equation}
where $\beta$ is the rotation angle of the second HWP. With
appropriate rotation angles $\alpha_{1}$, $\alpha_{2}$, and $\beta$,
we can realize arbitrary cylindrical vector beams represented by the
corresponding points ($\upsilon,\phi$) on the HOPS. This is the main
point of the theory.

\begin{figure}
\centerline{\includegraphics[width=8cm]{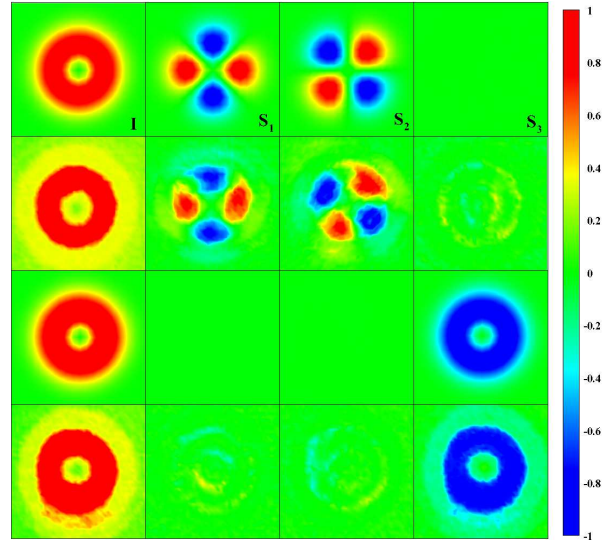}}
\caption{\label{Fig3} The Stokes parameters of the generated vector
beams corresponding to $A$ and $B$ on the first-order Poincar\'{e}
sphere (Fig.~\ref{Fig1}). The first column shows the intensity
distribution ($S_0$) and the next three columns are the Stokes
parameters $S_{1}$, $S_{2}$, and $S_{3}$, respectively. First and
third rows: the theoretical results for $A$ and $B$, respectively.
Second and fourth rows: the experimental results $A$ and $B$,
respectively.}
\end{figure}

To examine the polarization of the generated vector beams, we will
measure the Stokes parameters by a QP and a CCD camera (see
Fig.~\ref{Fig2}), $S_{1}$, $S_{2}$, and $S_{3}$, which are given by
\begin{equation}
S_{1}=\frac{I_{0^{\circ}}^{0^{\circ}}-I_{90^{\circ}}^{90^{\circ}}}{I_{0^{\circ}}^{0^{\circ}}+I_{90^{\circ}}^{90^{\circ}}},
S_{2}=\frac{I_{45^{\circ}}^{45^{\circ}}-I_{135^{\circ}}^{135^{\circ}}}{I_{45^{\circ}}^{45^{\circ}}+I_{135^{\circ}}^{135^{\circ}}},
S_{3}=\frac{I_{0^{\circ}}^{135^{\circ}}-I_{0^{\circ}}^{45^{\circ}}}{I_{0^{\circ}}^{135^{\circ}}+I_{0^{\circ}}^{45^{\circ}}}\label{asi},
\end{equation}
where $I_{j}^{i}$ stands for the intensity of the light recorded by
the CCD, and $i$ and $j$ are the optical axis directions of the QWP
and GLP with respect to the $x$ axis, respectively~\cite{Born1997}.
Note that the intensity profiles of $S_{3}$ depends on the polar
angle $\upsilon$. The linear polarized vector beams represented by
the equatorial points of the HOPS have $S_{3}={0}$ at each
transverse point. And $S_{3}={\pm1}$ correspond to circular
polarizations. But for the $S_{1}$ and $S_{2}$, the intensity
patterns are both dependent on the angles $\upsilon$ and $\phi$.

\begin{figure}
\centerline{\includegraphics[width=8cm]{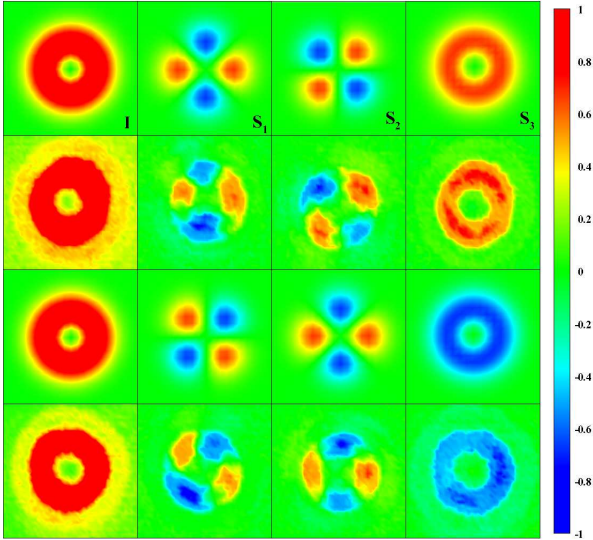}}
\caption{\label{Fig4}  The Stokes parameters of the generated vector
beams corresponding to $C$ and $D$ on the first-order Poincar\'{e}
sphere (Fig.~\ref{Fig1}). The first column shows the intensity
distribution and the next three columns are the Stokes parameters
$S_{1}$, $S_{2}$, and $S_{3}$, respectively. First and third rows:
the theoretical results for $C$ and $D$, respectively. Second and
fourth rows: the corresponding experimental results.}
\end{figure}

We first generate the vector beams on the equator and the south pole
($A$ and $B$) on the first-order Poincar\'{e} sphere as shown in
Fig.~\ref{Fig1}. For the vector beam on the point $A$
($\upsilon=\pi/2$ and $\phi=0$), from
Eqs.~(\ref{HPT})-(\ref{centroid}), we get $\alpha_{1}={90}$,
$\alpha_{2}={0}$, and $\beta={0}$. In order to achieve the vector
beam on the points B, we must ensure that $\alpha_{1}={0}$,
$\alpha_{2}={0}$, and $\beta$ be an arbitrary value. The Stokes
parameters of the generated beams are measured to verify the
theoretical prediction as shown in Fig.~\ref{Fig3}, which indicates
the high quality of the generated vector beams. Because the point
$A$ represents a linearly polarized vector beam, so $S_3=0$. While
$B$ is a circular polarized vortex beam, hence $S_1=S_2=0$ and
$S_3=-1$ with its intensity profile equals to the intensity of the
beam.

We now expect to obtain vector beams on the points $C$ with
$\upsilon=\pi/4$ and $\phi=0$, as well as $D$ with $\upsilon=3\pi/4$
and $\phi=\pi/2$ as shown in Fig.~\ref{Fig1}. This two kinds of
vector beams can be transformed from the beam on the point $A$.
According to Eqs.~(\ref{HPT}) and ~(\ref{centroid}), we can generate
the beam on the point $C$ by changing the angle $\alpha_{2}$ to
$49.9^{\circ}$. Similarly, the vector beam on the point $D$ can be
realized by rotating the PP1 with $\alpha_{1}=40.1^{\circ}$ and HH
with $\beta=22.5^{\circ}$, respectively. All the above
transformations to realize the vector beams on points $C$ and $D$
are schematically illustrated in Fig.~\ref{Fig1}. Then we measure
the Stokes parameters of the generated beams (Fig~\ref{Fig4}).
Remarkably, the profile of $S_1$ and $S_2$ consists of four
intensity lobes and the profile of $S_3$ exhibits a doughnut shape
with a dark center. The experimental results agree well with the
theoretical calculations. In order to obtain arbitrary cylindrical
vector beams on the first-order Poincar\'{e} sphere, we can adjust
the optical axis angles of PP1, PP2, and HH.

All the above discussions are limited to the case that the
topological charge is $\ell=+1$. Actually, we can also obtain the
cylindrical vector beams and realize its evolution on the HOPS by a
metasurface~\cite{Liu2014}, but it has less tunable to achieve a
cylindrical vector beams with other topological charges
(${\ell\neq{1}}$) than using the SLM. Here, this can be conveniently
achieved by modulating the phase picture displayed on the SLM, and
all the theoretical and experimental schemes are applicable in this
manipulation. To generate the vector beams on the other HOPS with
the opposite topological charge ${-\ell}$, one simple approach is to
invert the phase picture on the SLM, and another way is to insert a
HWP in each arm of the interferometer. Note that our scheme can also
be applied to generate the vector beams with more complex
polarization
distribution~\cite{Maurer2007,Wang2007,Chen2009,Cardano2013}.

In conclusion, we have developed a modified Mach-Zender
interferometer to generate arbitrary cylindrical vector beams on the
HOPS. By suitably modifying the magnitude and phase factors of the
interfering beams, we can achieve any desired vector beam on the
HOPS. The theoretical prediction is verified by measuring the
far-field Stokes parameters. Our scheme makes it possible to
generate versatile cylindrical vector beams with high tunability and
thereby providing a convenient way to evolve the polarization state
in any path on the high order Poincar\'{e} sphere.

This research was supported by the National Natural Science
Foundation of China (Grants Nos. 11274106, 61025024, and 11347120).

\end{document}